\documentclass[prd,aps,a4paper,eqsecnum,floatfix,nofootinbib,twocolumn,]{revtex4}  %

\newif\ifusesec
\usesectrue  
   
\usepackage{graphicx} 
\usepackage{amsmath,amsfonts,amssymb}
\usepackage{mathtools}
\usepackage{color}
\usepackage{slashed}

\makeatletter
\newsavebox{\@brx}
\newcommand{\llangle}[1][]{\savebox{\@brx}{\(\m@th{#1\langle}\)}%
  \mathopen{\copy\@brx\kern-0.5\wd\@brx\usebox{\@brx}}}
\newcommand{\rrangle}[1][]{\savebox{\@brx}{\(\m@th{#1\rangle}\)}%
  \mathclose{\copy\@brx\kern-0.5\wd\@brx\usebox{\@brx}}}
\makeatother

\newcommand{\beq}{\begin{equation}}
\newcommand{\eeq}{\end{equation}}
\newcommand{\bea}{\begin{eqnarray}}
\newcommand{\eea}{\end{eqnarray}}

\begin{document}

\title{Octupolar bremsstrahlung waveform up to the two-loop level \\ and the third-and-a-half post-Newtonian accuracy}

\author{Donato Bini$^{1}$, Thibault Damour$^{2}$, Andrea Geralico$^{1}$}  
  \affiliation{
$^1$Istituto per le Applicazioni del Calcolo ``M. Picone'' CNR, I-00185 Rome, Italy\\
$^2$Institut des Hautes Etudes Scientifiques, 91440 Bures-sur-Yvette, France\\
}

\date{\today}

\begin{abstract}
Extending our recent work (which focussed on the even-parity quadrupolar part of the waveform), we
 compute the  even-parity octupolar contribution, $U_3$, to  the gravitational waveform 
 $W \equiv \frac{c^4 r}{4G} \bar m^{i} \bar m^{j }  h_{i j}$ emitted during the scattering of two masses.
 We work within the Multipolar Post-Minkowskian (MPM) formalism, and use the 3.5 Post-Newtonian (PN) accurate radiation-reacted 
 quasi-Keplerian representation of the hyperbolic motion. We explicitly evaluate the frequency-domain value 
 $\hat U_3(\omega, \theta,\phi)$
 of  $U_3$ up to the 2-loop level, i.e. $ O(G^4)$ contributions to $h_{ij}(\omega, \theta,\phi)$, corresponding to $O(G^3)$ contributions to  $\hat U_3(\omega, \theta,\phi)$. As a crucial partial confirmation of our result, we find that the 
 1-loop truncation of our 3.5 PN frequency-domain MPM waveform  agrees with corresponding existing Effective Field Theory (EFT) results when  taking into account exactly the {\it same} (2.5PN-level) difference in the definitions of the 
 center-of-mass origins within the two formalisms that was deduced from our previous quadrupolar comparison.
\end{abstract}

\maketitle

\section{Introduction}

The gravitational radiation emitted during the scattering of two masses is currently
being extensively studied within several theoretical frameworks, and notably: the post-Newtonian-matched
Multipolar Post-Minkowskian (MPM) formalism,  various Effective-Field-Theory (EFT) approaches, and 
Self-Force theory. For recent examples of these various approaches, and entries into the original literature, see e.g.  Refs. \cite{Bini:2026dvn,Brunello:2025eso,Geralico:2026efi}.
All the approaches mentioned above are subject to theoretical limitations (especially when 
dealing with radiation-reaction effects) and face technical challenges. This makes it crucially important to
compare the results obtained by different formalisms both among themselves, and against the predictions of
general approaches to gravitational radiation, such as soft theorems (see, e.g., \cite{Sen:2024qzb}), and asymptotic frameworks (e.g., \cite{Boschetti:2026ogm,Compere:2026mdj}).

In a recent paper \cite{Bini:2026dvn} we have computed the even-parity quadrupolar part
of the  bremsstrahlung waveform at the two-loop level ($ h = O(G^4)$) and 3.5PN accuracy.
Here, we extend our previous results by computing the even-parity {\it octupolar bremsstrahlung radiation}.
We reach the same PM accuracy (i.e., two-loop level or $ h = O(G^4)$), and the same absolute 3.5PN accuracy on the
total complex waveform 
\beq \label{Wdef}
W(t_r,\theta,\phi) \equiv \lim_{r \to \infty} \frac{c^4 r}{4G} \bar m^{i} \bar m^{j }  h_{i j}(t_r, r,\theta,\phi), 
 \eeq
where 
\beq
{\bar {\bf m}}=\frac{1}{\sqrt{2}}\left(\frac{\partial \bf n}{\partial \theta} - \frac{i}{\sin \theta}  \frac{\partial \bf n}{\partial \phi}\right)\,,
\eeq
 is a null polarization vector
 orthogonal to the direction of emission ${\bf n}(\theta,\phi)$. However, an absolute 3.5PN accuracy on $W$
now corresponds to a 
fractional 3PN accuracy on the even-parity octupolar part, $U_3$, of $W$. We, indeed, recall that 
the full (time-domain) waveform at the absolute 3.5PN accuracy reads (using $\eta \equiv \frac{1}{c}$ as PN counting
parameter)
\bea
\label{W_deco}
W(t_r,\theta,\phi)&=& U_2+ \eta (V_2 +U_3) + \eta^2 (V_3+U_4)\nonumber\\ 
&+& \eta^3 (V_4+U_5)+ \eta^4 (V_5+U_6)\nonumber\\
&+& \eta^5 (V_6+U_7)+ \eta^6 (V_7+U_8)\nonumber\\
&+& \eta^7 (V_8+U_9)+O(\eta^8)\,.
\eea
Here $t_r$ denotes the retarded time, $U_l$ is the even-parity $l$-th radiative multipole, while 
$V_l$ is the odd-parity $l$-th radiative multipole. The polar angles $\theta, \phi$ are defined in the usual way, i.e.
such that the spatial direction of gravitational wave emission in the center-of-mass (cm) spatial frame
$e_x, e_y,e_z$ reads: $n^i(\theta,\phi) =[n_1,n_2,n_3]=[ \sin \theta \cos \phi, \sin \theta \sin \phi, \cos \theta]$.

Let us recall that (as in  our previous works) we use a cm spatial frame $e_x, e_y,e_z$ 
anchored to the averaged (conservative) momenta $\bar p_a$,
\beq
\label{bar_pa}
\bar p_a=\frac12 (p_a+p_a')\,,\qquad a=1,2\,,
\eeq
rather than to the incoming momenta $p_a$.
While the vector $e_x$ is aligned with the eikonal-type impact parameter, the vector $e_y$ is aligned with 
the spatial direction of $\bar p_1$  (i.e., the bisector between the incoming and the outgoing spatial momentum
of the first particle in the cm  frame), and the vector $e_z$ is orthogonal to the plane of motion, in the direction of the
(incoming) cm angular momentum.
The two spatial frames: $e_x, e_y,e_z$ (defined via $\bar p_a$) and the corresponding one $e_X,e_Y,e_Z$ (anchored on the incoming momenta  $p_a$) differ by a $O(G^1)$ rotation  involving  half  the (relative) conservative scattering angle,  $\chi_{\rm cons}/2$, around the $z$-axis  common to both frames: $e_Z=e_z$.

Our other notation and conventions are as follows: 
\begin{enumerate}
\item We use the mostly positive metric convention ($-+++$). 
\item We work within the MPM formalism, and express all multipole moments (source, gauge), considered
in the cm frame defined by the MPM approach (see above), in terms of the dynamical variables of the system by using  \lq\lq modified  harmonic coordinates," \cite{Blanchet:2013haa}. 
\item We consider a binary system of two point masses, $m_1$ and $m_2$, and denote:
$M=m_1+m_2$, $\mu =\frac{m_1m_2}{m_1+m_2}$, 
 $\nu\equiv \frac{\mu}{M} =\frac{m_1m_2}{(m_1+m_2)^2}$, and $X_a=\frac{m_a}{M}$ ($a=1,2$, with $X_1+X_2=1$,
 and $\nu = X_1 X_2$).   In the present work, we assume that $m_1 \geq m_2$, so that $X_1-X_2\equiv X_{12} = + \sqrt{1-4 \nu}$.  
This convention agrees with the one we used in Refs. \cite{Bini:2023fiz,Bini:2024ijq}, but differs from the one used in our
 recent work Ref. \cite{Bini:2026vaq}. 
\end{enumerate}

\section{MPM Waveform at 3.5PN: the octupolar contribution}

In order to compute the time-domain full waveform at the 3.5PN accuracy, Eq. \eqref{W_deco},
we need $U_2$ at 3.5PN accuracy, $V_2 ,U_3$ at 3PN accuracy, $V_3 , U_4$ at 2.5PN accuracy, $V_4 , U_5$ at 2PN accuracy, $V_5 ,U_6$ at 1.5PN accuracy, $V_7 ,U_8$ at Newtonian (N) accuracy, and $V_8 ,U_9$ at N accuracy. This task will be accomplished in successive steps.
In our previous work we computed $U_2$ at 3.5PN accuracy and at 2 loops, in this work we evaluate  $U_3$ at  3PN accuracy and at 2 loops. We recall that $U_3(t_r)=G^0+ G ^1+ G^2 + G^3$, where the $G^0$ contribution is constant in the time
domain, and does not appear in  $\hat U_3(\omega)$ when $\omega \neq 0$.
The PN-matched MPM formalism was constructed in Refs. 
\cite{Blanchet:1985sp,Blanchet:1986dk,Blanchet:1989ki,Damour:1990gj,Damour:1990ji,Blanchet:1998in,Poujade:2001ie}, and developed in many subsequent works by Blanchet et al. (see \cite{Blanchet:2013haa} for a review).
Here, we will particularly make use of the results of  Faye, Blanchet and Iyer \cite{Faye:2014fra} who derived the relation
between the octupolar radiative moment and the source.

Using Ref. \cite{Faye:2014fra},  the relation between the radiative octupole $U_{ijk}$ (
entering the waveform)
 and the $M_{ijk}$ 
 canonical octupole (parametrizing the external MPM metric) is given by
\beq	 
\label{UvsM}
U_{ijk} =  M_{ijk}^{(3)} +  U_{ijk}^\text{1.5PN}{}  +  U_{ijk}^\text{2.5PN} +U_{ijk}^\text{3PN}{}_{\rm  tail(tail)}\,, 
\eeq
where the second contribution on the right hand side (rhs), which is the standard linear tail, contains an overall factor 
$\eta^3= \frac1{c^3}$  and reads,
\bea
\label{U3tail}
U_{ijk}^\text{1.5PN} 
&= &
 2 G {\mathcal M}\eta^3 \int_0^{+\infty} d\tau\,  M_{ijk}^{(5)}(t_r-\tau)\, \ln \left(\frac{\tau}{C_{I_3}}\right) 
,\qquad
\eea
with
\beq
C_{I_3}=\frac{2 b_0}{c}e^{-\frac{97}{60}}\,,\qquad {\mathcal M}=\frac{E}{c^2}\,,
\eeq
while the third contribution  (denoted $ U_{ijk}^\text{2.5PN}$) contains an overall factor $\eta^5= \frac1{c^5}$ 
and can be conveniently decomposed in various terms keyed by the multipoles they contain:
\bea
U_{ijk}^\text{2.5PN}&=& U_{ijk}^\text{2.5PN (mem)}+ U_{ijk}^{ {\rm 2.5PN }, I_2I_3}\nonumber\\
&+& U_{ijk}^{ {\rm 2.5PN }, J J_2}+ U_{ijk}^{ {\rm 2.5PN }, I_2J_2}\,.
\eea
The various building blocks of  $U_{ijk}^\text{2.5PN}$ 
are summarized in Table \ref{table:1}. 

Finally, the last contribution is the $O(\eta^6)$  tail of tail term given by
\beq
\label{U3tailoftail}
U_{ijk}^\text{3PN}{}_{\rm  tail(tail)}=  2 G^2 {\mathcal M}^2\eta^6 \!\!\int_{0}^{+\infty} \!\!\!d\tau\, 
M_{ijk}^{(6)}(t_r-\tau)\,{\mathcal L}_2(\tau)\,,
\eeq
where
\bea
{\mathcal L}_2(\tau)&=&\ln^2\left(\frac{c \tau}{2b_0}\right) + \frac{97}{30} \ln\left(\frac{c \tau}{2b_0}\right)\nonumber\\
&-& \frac{13}{21} \ln\left(\frac{c \tau}{2r_0}\right)
+ \frac{13283}{88200}\nonumber\\
&=&\ln^2\left(\frac{\tau}{C_{I_3}}\right) - \frac{13}{21} \ln\left(\frac{\tau}{C_*}\right)\,,
\eea
with
\beq
C_*=2r_0 e^{-\frac{65127}{36400}}\,.
\eeq
Note that the tail-of-tail term involves two different length scales: $r_0$ and $b_0$.
%
%
\begin{table*}
\caption{\label{table:1} Building blocks for the explicit computation of $U_{ijk}^\text{2.5PN}$  from 
Ref. \cite{Faye:2014fra}. 
The underlined indices within angled brackets are excluded from the STF projection.
Beware of  three sign misprints in Eq. (248b) of Ref. \cite{Blanchet:2013haa}  affecting the terms proportional to $M_2 S_2^{(5)}$, $M_2^{(1)} S_2^{(4)}$ and $M_2^{(2)} S_2^{(3)}$
in the $U_{ijk}^{ {\rm 2.5PN  }, I_2J_2}$ line below. 
}
\begin{ruledtabular}
\begin{tabular}{l|l}
$U_{ijk}^\text{2.5PN (mem)}$ &$\frac{G}{c^5}  
		\int_0^{+\infty} \! d\tau\left[-\frac{1}{3}M^{(3)}_{a\langle i}M^{(4)}_{jk\rangle a}- \frac{4}{5}\epsilon_{ab\langle i}M^{(3)}_{j\underline{a}}S^{(3)}_{k\rangle b}\right]\!(t_r-\tau) $\\
$U_{ijk}^{ {\rm 2.5PN }, I_2I_3}$& $\frac{G}{c^5} \Biggl[
		\frac{1}{4} M^{}_{a\langle i} M^{(6)}_{jk\rangle a}
		+ \frac{1}{4} M^{(1)}_{a\langle i} M^{(5)}_{jk\rangle a}
		+ \frac{1}{4} M^{(2)}_{a\langle i} M^{(4)}_{jk\rangle a}
		- \frac{4}{3} M^{(3)}_{a\langle i} M^{(3)}_{jk\rangle a}- \frac{9}{4} M^{(4)}_{a\langle i} M^{(2)}_{jk\rangle a}- \frac{3}{4} M^{(5)}_{a\langle i} M^{(1)}_{jk\rangle a}+ \frac{1}{12} M^{(6)}_{a\langle i} M^{}_{jk\rangle a}\Biggr]$\\
$U_{ijk}^{ {\rm 2.5PN }, J J_2}$& $ \frac{12}{5}\frac{G}{c^5}  S^{}_{\langle i} S^{(4)}_{jk\rangle}$\\
$U_{ijk}^{ {\rm 2.5PN  }, I_2J_2}$&$  \frac{G}{c^5}   \epsilon_{ab\langle i}\bigg[
		-\frac{9}{5} M^{}_{j\underline{a}}S^{(5)}_{k\rangle b}
		- \frac{27}{5} M^{(1)}_{j\underline{a}}S^{(4)}_{k\rangle b}
		- \frac{8}{5} M^{(2)}_{j\underline{a}}S^{(3)}_{k\rangle b}
		+ \frac{12}{5} M^{(3)}_{j\underline{a}}S^{(2)}_{k\rangle b}
		+ \frac{3}{5} M^{(4)}_{j\underline{a}}S^{(1)}_{k\rangle b}
		+ \frac{1}{5} M^{(5)}_{j\underline{a}}S^{}_{k\rangle b}
		\bigg]$\\
$U_{ijk}^{ {\rm 2.5PN  }, I_3J_1}$&$ \frac{G}{c^5} \frac{9}{20}  \epsilon_{ab\langle i}  M^{(5)}_{jk\rangle a}S^{}_b$\\
\end{tabular}
\end{ruledtabular}
\end{table*}

In addition to the relation  \eqref{UvsM} between $U_{ijk}$ and the canonical moments, we also need the expression
of the canonical octupole in terms of source and gauge moments, which is given by
\bea  
\label{M3}
M_{ijk} &=&  I_{ijk} +  4G\eta^5 \biggl[ W^{(2)} I_{ijk}- W^{(1)} I_{ijk}^{(1)}+3\, I_{\langle ij} Y_{k\rangle }^{(1)}\biggr]\nonumber\\ 
&+&O\left(\eta^7\right)\label{M3} \,,\nonumber\\ 
&\equiv& I_{ijk} + M_{ijk}^{WI_3}+  M_{ijk}^{I_2Y_1}+O\left(\eta^7\right)\,.
\eea
The explicit expressions (at the Newtonian approximation)
of the gauge moments, $W$ and $Y_k$ can be found, e.g., in Table I of our previous paper \cite{Bini:2026dvn}.

Finally, we need to insert in Eq. \eqref{M3} the 3PN-accurate expression  of the  source octupole  $I_{ijk}$, which includes
a crucial (time-odd)   2.5PN contribution, say $I_{ijk}^{\eta^5}$.
These terms have been computed in Eqs. (4.9)-(4.10) of Ref. \cite{Faye:2014fra}  (see also  Eqs. (3.6)-(3.7)) of
 Ref. \cite{Henry:2021cek}). They read
\bea
I_{ijk}&=&I_{ijk}^{\eta^0} + I_{ijk}^{\eta^2}+ I_{ijk}^{\eta^4}+I_{ijk}^{\eta^5} + I_{ijk}^{\eta^6}\,,
\eea
where we note that the time-odd piece $I^{\eta^5}_{ijk}$ explicitly reads
\bea
I^{\eta^5}_{ijk}
&=& -\frac{\nu^2}{c^5} (m_1-m_2)\left[-\frac{56}{9 } \left(\frac{GM}{r}\right)^2\dot r   x^{\langle ijk \rangle}\right.\nonumber\\
&+&\left.
r \left(\frac{232}{15}\left(\frac{GM}{r}\right)^2 - \frac{12}{5} \frac{GM}{r} v^2  \right) v^{\langle i}x^{jk\rangle}\right]\,.\nonumber\\
\eea
Here, $x^i\equiv x^i_{12}= x^i_1-x^i_2$ and $v^i\equiv v^i_{12}=\dot x_{12}^i$ are the relative positions and velocities of the binary system in the cm frame. 

In the incoming state, the $\eta^5$ contribution to the octupole moment asymptotes to
\beq
I^{\eta^5}_{ijk} \approx + \frac{12}{5}  \frac{\nu^2 G M }{c^5} (m_1-m_2) v^2  v^{\langle i}x^{jk\rangle}\,.
\eeq
This contribution corresponds to the following $O(G \eta^5)$  shift of the 
position of the MPM center of mass:
\beq
\label{Xieta5}
X^i_{\eta^5}= +  \frac{4}{5}  \frac{G (m_1-m_2) \nu}{c^5} (v^2  v^{ i})_{t\to-\infty}\,.
\eeq
This shift of the position of the center of mass will be further discussed in the next section.

\section{Evaluation of the orbit}

The source (and gauge) multipole moments are originally computed in terms of the positions and velocities of the two particles $x_1^i(t)$ and $x_2^i(t)$ 
(e.g., $I_{ijk}(t)^{\eta^0}= m_1 x_1^{\langle ijk \rangle}+ m_2 x_2^{\langle ijk \rangle}$). 
We then need explicit expressions for the two absolute (radiation-reacted) 
hyperboliclike motions $x_1^i(t)$ and $x_2^i(t)$. The first step in computing such  expressions  consists
in writing the two absolute motions, $x_1^i(t)$ and $x_2^i(t)$, 
in terms of the relative motion, $x^i(t)=x_1^i(t)- x_2^i(t)$, by using
 an explicit form of the variation of the \lq\lq center-of-mass constant" of the binary system, 
$K^i=G^i - t P^i$, together with an explicit form of the variation of the total linear momentum of the binary system, $P^i$. 
When working, as we do, at the 3.5PN accuracy, this problem is facilitated by two facts: (i) the radiative losses of $P^i$ and $K^i$ start at the 3.5PN level; and
(ii) the 3.5PN-level motion of the center of mass of the system can be neglected when evaluating the 3.5PN-accurate waveform in the (incoming) cm system. Indeed, it would only be needed for the radiative quadrupole, but it happens
to have a negligible contribution to $I_{ij}$. This follows because of the transformation property of a 
Newtonian quadrupole under translation of the origin away from the cm: indeed, $x_a^i=X^i +x_{a \, \rm cm}^i $ implies $I_{ij}= I_{ij}^{\rm cm} + M X^{\langle ij \rangle} + O(\eta^2)$. [See, \cite{Damour:1990ji} for the relativistic law
of variation of the source multipoles under a translation of the spatial origin.]

When neglecting the 3.5PN level radiative losses of $P^i$ and $K^i$, i.e., when working at the 3PN level (but including
the leading-order radiation reaction at 2.5PN) we have {\it conservation laws} for $P^i$ and $K^i= G^i- t P^i$, with
\bea
P_i= P_i^{\eta^0}+ P_i^{\eta^2}+P_i^{\eta^4}+P_i^{\eta^5}+P_i^{\eta^6}, \nonumber\\ 
G_i= G_i^{\eta^0}+ G_i^{\eta^2}+G_i^{\eta^4}+G_i^{\eta^5}+G_i^{\eta^6}\,,
\eea
where the quantities on the right-hand side are functions of positions and velocities (or momenta), e.g.,
at the Newtonian level: $P_i^{\eta^0}= m_1 v_1^i + m_2 v_2^i $ and $G_i^{\eta^0}= m_1 x_1^i + m_2 x_2^i $.
The aim of these reminders is to emphasize that the 2.5PN-accurate definition of the cm, namely $P_i^{\rm cm}=0$
and $K_i^{\rm cm}=0$, which implies  $G_i^{\rm cm}=0$ includes a time-odd radiation-reaction contribution, namely
(denoting $G_i^{\rm 3PN, cons} \equiv G_i^{\eta^0}+ G_i^{\eta^2}+G_i^{\eta^4}+G_i^{\eta^6}$)
\beq
G_i^{\rm 3PN, cons} + G_i^{\eta^5}=0\,,
\eeq
where the 2.5PN contribution to the cm position vector (first derived in Refs.  \cite{Damour:1982wm,Damour:1983tz} from the 2.5PN equations of motion; and reobtained within the MPM formalism in \cite{Blanchet:1996wx})) is equal  to (the last expression being in the cm frame)
\bea \label{Gieta5}
G_i^{\eta^5}&=& \frac45 G m_1 m_2 \eta^5 \left( v_{12}^2- 2 \frac{G (m_1+m_2)}{r_{12}} \right) \left(v_1^i+ v_2^i \right) \nonumber\\
&=& -  \frac45 G m_1 m_2 \eta^5 X_{12}  \left( v_{12}^2- 2 \frac{G (m_1+m_2)}{r_{12}} \right) v_{12}^i
\,.\nonumber\\
\eea
This expression has three important properties within our present context: (i) it is of order $G^1$, i.e. 1PM; (ii) it does not vanish
in the asymptotic scattering states; and (iii) the corresponding  asymptotic incoming time-odd contribution to the position of the cm, namely
\bea
\label{Xieta5cm}
X^i_{\eta^5}&\equiv& -\frac{G^i_{\eta^5}}{M}\big|_{t\to -\infty}\nonumber\\
&=& +\frac{4}{5}  \frac{G (m_1-m_2) \nu}{c^5} (v_{12}^2  v_{12}^{ i})_{t\to-\infty}\,,
\eea
coincides both with the shift deduced in Eq. \eqref{Xieta5} from the time-odd contribution to the octupole moment, and with the  leading PN-order value of  
the vector $\beta_i^{\rm VV}$ describing the dipolar part of the Veneziano-Vilkovisky supertranslation
\cite{Veneziano:2022zwh}
\beq
\left[\beta^{\rm VV}({\bf n}) \right]^{\rm dipole}= \beta_i^{\rm VV} n^i\,.
\eeq
See Ref. \cite{Bini:2026dvn} for the full multipolar expansion of the Veneziano-Vilkovisky supertranslation.

In  previous works \cite{Bini:2025rng,Bini:2026suo} we have computed (at the 3.5PN accuracy) 
the radiation-reaction contributions to the relative hyperboliclike motion $x^i(t)=x_1^i(t)- x_2^i(t)$ by using harmonic coordinates,
and   a quasi-Keplerian parametrization for the orbit.
The final solution is expressed in terms of the eccentric anomaly and the unperturbed (incoming) values of the semilatus rectum and radial eccentricity.
The  radiation-reacted part of the relative motion (truncated at $O(G^3)$ for simplicity) is conveniently expressed 
in terms of the rescaled time variable $T\equiv \frac{ p_\infty t}{b }$ and of the gauge-invariant parameters $p_\infty
\equiv \sqrt{\gamma^2-1}$  and $b$ (impact parameter in the incoming cm).

Here we do not need the full 3.5PN accuracy on the relative hyperbolic motion, but only the 3PN one (with inclusion
of the 2.5PN radiation-reaction contribution). We can therefore distinguish a 3PN-accurate conservative part (fully represented in a quasi-Keplerian form) and a 2.5PN radiation-reaction part,
say  $x^i(t)=x^i_{\rm 3PN \, cons}(t)+\delta^{\rm rr}x^i(t)$. Truncating to the $G^2$ and $G^3$ terms, the
2.5PN radiation-reaction contribution to the motion has the structure:
\bea
\delta^{\rm rr}x^i(t)&=& \frac{G^2 M^2 \nu p_\infty \eta^5}{b} \delta^{\rm rr,G^2} x^i(T)\nonumber\\ 
&+& \frac{G^3 M^3 \nu \eta^5}{p_\infty b^2} \delta^{\rm rr,G^3} x^i(T)\,.\nonumber
\eea
Let us also recall that while $x^i_{\rm 3PN \, cons}(t)$ is time-symmetric (and is naturally projected along 
the corresponding conservative frame $e_x, e_y, e_z$ based on averaged momenta), the radiation-reaction part
$\delta^{\rm rr}x^i(t)$ is time-asymmetric, and is defined by using the boundary condition 
 $\lim_{t \to - \infty} \delta^{\rm rr}x^i(t) = 0$ (which specifies the initial state of the system).

After having derived explicit expressions for the time-domain radiative octupole moment $U_{ijk}(t)$ we computed its
frequency-domain version:
\beq
\hat U_{ijk}(\omega)= \int_{- \infty}^{+ \infty} dt_r e^{i \omega t_r} U_{ijk}(t_r)\,.
\eeq
As usual, we will only consider the positive frequency axis, $\omega >0$. Indeed, the real character of
the time-domain octupole $U_{ijk}(t_r)$ implies that its Fourier transform satisfies the reality
conditions $\left(\hat U_{ijk}(\omega)\right)^*= \hat U_{ijk}(-\omega)$ relating $\omega >0$ to $\omega <0$.
The positive frequency axis therefore contains the full information about $\hat U_{ijk}(\omega)$. 

\section{Explicit evaluation of $U_3$}

Let us consider the octupolar contribution to the (helicity minus 2) complex waveform $W$, Eq. \eqref{Wdef}, namely
\beq
U_3=\frac1{3!}U_{ijk}n^i \bar m^j \bar m^k\,,
\eeq
computed (both in the time  and frequency domains) along hyperboliclike orbits and at the 3PN and $O(G^3)$ accuracy.
We distinguish:
\begin{enumerate}
\item $U_3^{I_3}$, to be evaluated at the 3PN fractional accuracy;
\item $U_{3}^{M_{3}^{WI_3}}$ and $U_{3}^{M_3^{I_2Y_1}}$, to be evaluated at the N level;
\item $U_{3}^\text{1.5PN}$, to be evaluated at the 1PN fractional accuracy;
\item $U_{3}^\text{2.5PN}$ and $U_{ijk}^\text{3PN}{}_{\rm  tail(tail)}$, to be evaluated at the N level,
\end{enumerate}
so that
\bea
U_3&=&\underbrace{U_3^{I_3}}_{\rm 3PN}+\underbrace{U_{3}^{M_{3}^{WI_3}}+U_{3}^{M_3^{I_2Y_1}}}_{\rm N}+\underbrace{U_{3}^\text{1.5PN}}_{\rm 1PN}\nonumber\\
&+&
\underbrace{U_{3}^\text{2.5PN}+U_3^\text{3PN}{}_{\rm  tail(tail)}}_{\rm N}\,.
\eea

The easiest part of the computation concerns the hereditary contribution $U_3^\text{1.5PN}$ and $U_3^\text{3PN}{}_{\rm  tail(tail)}$, which involve logarithmic integrals of the general form (see e.g., Ref. \cite{Bini:2021qvf})
\beq
A_m(\omega, C_X)= \int_0^{+\infty}  d\tau\,  e^{ i\omega \tau }  \ln^m\left(\frac{\tau}{C_{X}} \right)\,,
\eeq
that is, for $m=1,2$, 
\bea
A_1(\omega, C_X)
&=& -\frac{\pi}{2|\omega|}-\frac{i}{\omega}\ln (C_X |\omega|e^\gamma)\,,\nonumber\\
A_2(\omega, C_X)&=& \frac{\pi}{|\omega|}\ln (C_X |\omega|e^\gamma)\nonumber\\
&+& \frac{i}{\omega}\left[ \ln^2 (C_X |\omega|e^\gamma)-\frac{\pi^2}{12}\right]\,.
\eea
The Fourier transform of the tail term, Eq. \eqref{U3tail}, then reads
\beq
\hat U_{3}^\text{1.5PN}(\omega) = 2 G {\mathcal M}\eta^3 (-i\omega)^2\hat U_3^{I_3}(\omega) A_1(\omega,C_{I_3})\,,
\eeq
where it is enough to insert the 1PN-accurate values of both ${\mathcal M}=\frac{E}{c^2}$ and $\hat U_3^{I_3}(\omega)$. 
The tail-of-tail term, Eq. \eqref{U3tailoftail}, instead, turns out to be
\bea
\hat U_{3}^\text{3PN}{}_{\rm  tail(tail)}(\omega)&=&
 2 G^2 {\mathcal M}^2\eta^6 (-i\omega)^3\hat U_3^{I_3}(\omega)\nonumber\\
&\times&
\left[A_2(\omega,C_{I_3})-\frac{13}{21}A_1(\omega,C_*)\right]\,.\nonumber\\
\eea

The contributions $U_{3}^{M_{3}^{WI_3}}$ and $U_{3}^{M_3^{I_2Y_1}}$ involving gauge moments are $O(\eta^5)$, so that they have to be evaluated at the Newtonian level, and then Fourier-transformed.
Similarly, $U_{3}^\text{2.5PN}$ is made of five terms (listed in Table \ref{table:1}), to be evaluated at the Newtonian order: memory (2 terms), $I_2I_3$ (7 terms), $J J_2$ (1 term), $I_2J_2$ (6 terms), $I_3J_1$ (1 term),
\bea
U_{3}^\text{2.5PN}&=& U_{3}^\text{2.5PN (mem)}+ U_{3}^{ {\rm 2.5PN }, I_2I_3}+ U_{3}^{ {\rm 2.5PN }, J J_2}\nonumber\\
&+&
U_{3}^{ {\rm 2.5PN }, I_2J_2}+U_{3}^{ {\rm 2.5PN }, I_3J_1}\,.
\eea
The only contribution to $U_3$ to be evaluated at the 3PN accuracy is $U_3^{I_3}$, coming from $I_{ijk}^{(3)}$.
The Fourier transform of these instantaneous terms is done in terms of Bessel functions as well as other special functions that we will briefly recall below.

We conveniently express our final Fourier space results in terms of the dimensionless frequency-related variable
\beq
u \equiv \frac{\omega b}{p_\infty},
\eeq
which is dual to the dimensionless rescaled time variable $T\equiv \frac{ p_\infty t}{b }$
(and which should not be confused with a usual notation for the retarded time).
For the reasons mentioned above it is enough to consider the positive $u$ axis.

The final result is then expanded in powers of $G$, 
\bea
\hat U_3(u)&=&  \hat U_3^{G^1}(u)+ \hat U_3^{G^2}(u)+\hat U_3^{G^3}(u)\,,
\eea
where each PM term $\hat U_3^{G^n}(u) \propto G^n$  is PN expanded (i.e., expanded in powers of $\eta p_\infty$) up to $O(\eta^6)$ included. 

In the time domain $U_3(t_r)$ can be written as a polynomial in the equatorial components (indices $x,y$, or $1,2$)
of $n^i$ and $\bar m^i$, i.e. we have the structure (with $a\in \{x,y\} \equiv \{1,2\}$)
\beq
U_3(t_r)= \sum_{a,b,c=1,2} C_{abc}(t_r) n^a {\bar m}^b {\bar m}^c\,,
\eeq
where the coefficients $C_{abc}(t_r)$ are real. When going to the frequency domain (and using the dimensionless frequency
variable $u$) we have
\beq
\hat U_3(u)= \sum_{a,b,c=1,2} \hat C_{abc}(u) n^a {\bar m}^b {\bar m}^c\,,
\eeq
where the Fourier-transformed coefficients $\hat C_{abc}(u)$ are complex but satisfy the reality condition
\beq
\hat C_{abc}(-u)= [\hat C_{abc}(u)]^*\,.
\eeq
As already mentioned above these reality conditions allows us to restrict $u = \omega b/p_\infty$ to the positive axis $u>0$.
Let us also note the scaling of the PN expansion of the various PM contributions, $G^n$, with $n=1,2,3$, to $\hat C_{abc}(u)$:
\bea
\hat C_{abc}(u)^{G^n} &\sim& GM^2 \nu X_{12}\left(\frac{GM}{b p_\infty^2} \right)^{n-1} \nonumber\\
&\times&[c_{0}+c_2\eta^2 p_\infty^2 + c_3\eta^3 p_\infty^3+c_4\eta^4 p_\infty^4\nonumber\\
&&
+c_5 \eta^5 p_\infty^5  +c_6 \eta^6 p_\infty^6  + O(\eta^7) ]\,,
\eea
where the coefficients on the rhs only depend on $u$, besides the standard PN-type dependence on $\nu$.
At each PM order $n$ the PN coefficients $c_k$ are polynomials in $\nu$, say  $P_N(\nu)$, with an order $N$ increasing with the PN order: $c_0\sim P_0(\nu)$, $c_2\sim P_1(\nu)$,  $c_3\sim P_0(\nu)$ (for $n>1$), $c_4\sim P_2(\nu)$,  $c_5\sim P_1(\nu)$ (for $n>1$), and  $c_6\sim P_3(\nu)$.  

One should recall (see e.g., \cite{Bini:2023fiz,Bini:2024rsy,Bini:2024ijq,Bini:2026dvn})
the various special functions of $u$ which enter at each PM order: besides $K_0(u)$, $K_1(u)$ and $e^{-u}$ which appear starting at  $O(G^1)$, we have at $O(G^3)$
\bea
Q^{\rm as}_1(u)&=& \int_{-\infty}^\infty dT \frac{e^{iuT}{\rm arcsinh}(T)}{1+T^2}\,,\nonumber\\
Q^{\rm at}_{\frac12} (u)&=& \int_{-\infty}^\infty dT \frac{e^{iuT}{\rm arctan}(T)}{(1+T^2)^{1/2}}\,, \nonumber\\
Q^{\rm as2}_1(u)&=& \int_{-\infty}^\infty dT \frac{e^{iuT}{\rm arcsinh}^2(T)}{1+T^2}\,,
\eea
for which analytic closed forms exist in terms of second derivatives of the Bessel K functions with respect to the order, and in terms of Meijer G functions.
Our results (including the explicit expressions for the above integrals) are not displayed here but are available in the associated ancillary file 
\cite{anc}.
For example, the LO results for $U_3^{G^3}$ read
\begin{widetext}
\bea
\hat U_3^{G^3}(u) &=& \frac{G^3M^4}{p_\infty^4} \frac{\nu X_{12}}{b^2}\left[A_0 K_0(u)+A_1 K_1(u)+A_2Q^{\rm as2}_{\frac12}(u) +A_3 \frac{d}{du}Q^{\rm as2}_{\frac12}(u)\right]
+O\left(\frac1{p_\infty^2}\right)\,,
\eea
with
\bea
A_0&=&  \bar m_1^2 \left[\left(\frac{1}{2} u^3+3  u\right)i n_1+\left(-\frac{13}{6} u^2-2 \right)n_2 \right] 
+\bar m_1 \bar m_2  \left[ \left(-\frac{13}{3} u^2-4 \right)n_1 -i \left( u^3+\frac{29}{3} u\right) n_2 \right]\nonumber\\ 
&+& 
  \bar m_2^2  \left[ \left(-\frac{1}{2} u^3-\frac{29}{6}u\right) i n_1+3\left(u^2+1 \right)n_2\right]\,,    \nonumber\\
A_1&=&  \bar m_1^2 \left[\left(\frac{3}{2}  u^2+2\right)i n_1+\left(-\frac{1}{2}u^3-\frac{11}{3}u\right)n_2   \right]
+\bar m_1 \bar m_2 \left[\left(- u^3-\frac{22}{3} u\right)n_1+\left(-\frac{14}{3} u^2-6 \right) i n_2 \right]\nonumber\\
&+& \bar m_2^2 \left[ \left( -\frac{7}{3}  u^2-3\right)i n_1+\left(\frac{1}{2}u^3+\frac{11}{2} u\right)n_2 \right] \,, \nonumber\\
A_2&=&  \bar m_1^2 \left( -\frac{1}{4}n_2 u^4 \right) 
+\bar m_1 \bar m_2 \left[ -\frac{1}{2}n_1 u^4-\frac{5}{6} i n_2 u^3\right]
+\bar m_2^2 \left[ -\frac{5}{12} i n_1 u^3+\left(\frac{1}{4} u^4+\frac{1}{2}u^2\right)n_2\right]\,, \nonumber\\ 
A_3&=&  \bar m_1^2 \left(  \frac{1}{3}n_2 u^3-\frac{1}{4} i n_1 u^4\right)+ 
\bar m_1 \bar m_2 \left[\frac{2 }{3}n_1 u^3+\left(\frac{1}{2} u^4+u^2\right)  i n_2\right]
+ \bar m_2^2 \left[ \left(\frac{1}{4}  u^4+\frac{1}{2}u^2\right)i n_1-\frac{3}{4} n_2 u^3\right]  
\,.\nonumber\\
\eea
\end{widetext}
[Here, the components $n_3$ and $(\bar m_3)^2$ have been eliminated by using $n\cdot \bar m=0$ and $\bar m\cdot \bar m=0$.]

In view of the numerous intermediate building blocks that went into the computation of our  octupolar frequency-domain waveform,
we felt important to perform several checks of our final results. 
Similarly to our previous work on the quadrupolar waveform \cite{Bini:2026dvn}
we performed five different checks: (i) the disappearance of  the MPM related UV-regulator scale $r_0$ (which cancels against
the scale dependence of the source octupole moment $I_{ijk}$ \cite{Blanchet:1997jj,Goldberger:2009qd,Faye:2014fra});
(ii) the factorization of the gauge scale $b_0$  as $\propto e^{\frac{i2G{\mathcal M}\omega}{c^3}\ln b_0}$;
(iii)  the agreement of the extreme mass-ratio limit $\nu \ll 1$ against the recent results of \cite{Geralico:2026efi};
(iv) compatibility with the existing high-PN accuracy expansion of the 1-loop  $h_c=O(G^3)$, i.e.,  $W=O(G^2)$  waveform \cite{Brandhuber:2023hhy,Herderschee:2023fxh,Georgoudis:2023lgf,Heissenberg:2025fcr}; and finally
 (v) the soft limit, $\omega \to 0$ (i.e. $u \to 0$).
All our checks have been successful. Let us only give below a few details of the last two checks. 

\section{Comparison with amplitude-based EFT waveform results at $O(G^2)$ (1-loop)}

Starting from the one-loop EFT waveform of Refs. \cite{Brandhuber:2023hhy,Herderschee:2023fxh,Georgoudis:2023lgf,Bohnenblust:2023qmy} and its accurate PN expansion derived in Ref. \cite{Heissenberg:2025fcr}
we evaluated the difference $\delta \hat U_3(\omega)=U_3^{\rm EFT}-U_3^{\rm MPM}$  at $O(G^2)$,
and at our 3PN accuracy.  
All the PN terms in the difference cancel except for the ones  at  the fractional 2.5PN level, containing an overall factor $\nu^2$, which are proportional to $\nu^2 \eta^5 p_\infty^5/p_\infty^2=\nu^2 \eta^5 p_\infty^3$.
Explicitly, we found for the nonvanishing ($s=-2$) spin-weighted spherical harmonics of $\delta \hat U_3(\omega)$
\begin{widetext}
\bea
\delta \hat U_{33}&=&  \nu^2 X_{12}\frac{G^2 M^3 p_\infty^3}{b} \frac{8}{5}\sqrt{\frac{2 \pi }{21}}  u \left[\left(u+\frac{1}{2}\right) K_0(u)+(u+1) K_1(u)\right]
\,,\nonumber\\
\delta \hat  U_{31}&=&\nu^2 X_{12}  \frac{G^2 M^3 p_\infty^3}{b} \frac{8}{15}\sqrt{\frac{2 \pi }{35}}  u \left[\left(u+\frac{3}{2}\right) K_0(u)+(u+1) K_1(u)\right]
\,,\nonumber\\
\delta \hat  U_{3\bar 1}&=& -\nu ^2 X_{12} \frac{G^2 M^3 p_\infty^3}{b}\frac{8}{15} \sqrt{\frac{2 \pi }{35}} u \left[\left(u-\frac{3}{2}\right) K_0(u)+(1-u) K_1(u)\right]
\,, \nonumber\\
\delta \hat  U_{3\bar 3}&=& -\nu ^2  X_{12} \frac{G^2 M^3 p_\infty^3}{b}\frac{8}{5} \sqrt{\frac{2 \pi }{21}} u \left[\left(u-\frac{1}{2}\right) K_0(u)+(1-u) K_1(u)\right]
\,.
\eea
\end{widetext}

However, this difference is not a disagreement between EFT and MPM results. 
Indeed, it was found in our previous 3.5PN-accuracy quadrupolar EFT-MPM comparison \cite{Bini:2026dvn} that there
was a mismatch of the form
\beq \label{cyW}
\delta  W=i  \eta \omega (c_y n^y )W^{\rm tree}\,,
\eeq
where $n^y=\sin\theta \sin\phi$ is the $y$ component of the emission direction ${\mathbf n}$ and
\beq \label{cy}
c_y=+\frac45 \frac{G  \eta^5  m_1m_2 (m_1 - m_2) }{(m_1 + m_2)^2}  p_\infty^3+O(p_\infty^5)\,.
\eeq
[Here, we define $c_y$ so that it has dimensions of a length.] 
Such a mismatch only corresponds to a difference in the choice of the spatial cm origin in the two considered waveforms. 
The important point, is that, in our previous comparison, we had a
 $O(\eta^7)$ (3.5PN-level) mismatch in the quadrupolar part of $W$ coming from combining the $O(\eta^6 G^1)$ $\eta\omega c_y$ shift,  Eq. \eqref{cyW}, with the octupolar  part, $\eta U_3^G$ of $W_{\rm tree}$. Here, we have  a $O(\eta^6)$ mismatch in $W^{G^2 \, \rm even}_3=\eta U_3^{G^2 \, \rm even}$ corresponding to a $O(\eta^5)$ (2.5PN-level) mismatch in $U_3$. 
 However, we have checked that we have simply
 \beq
\delta  \hat U_3=\left[i\omega (c_y n^y )W^{\rm tree}\right]_{l=3^+}\,,
\eeq
with the {\it same} cm shift $c_y$, Eq. \eqref{cy}, that we had found in our previous quadrupolar comparison.
This is a non trivial confirmation of our new octupolar results.

Let us finally comment that the physical origin of the difference in the definitions of the position of the center of mass
(or more precisely of the location of the cm worldline) in the two formalisms (MPM and EFT) is unclear. We conjectured
in \cite{Bini:2026dvn} that it might be due to having included in the EFT waveform the full Veneziano-Vilkovisky supertranslation instead of its $l \geq 2$ projection. 
Our present result is compatible with this interpretation. We wish, however, to point out that, as said above, see Eq. \eqref{Xieta5cm},   
 the PN definition of the {\it incoming} MPM cm position $X^i$  contains a time-odd contribution equal to 
\beq
X^i_{\eta^5}=
 +\frac{4}{5}  \frac{G (m_1-m_2) \nu}{c^5} (v_{12}^2  v_{12}^{ i})_{t\to-\infty}=+c_y e_y^i\,.
\eeq
[Here the last rhs should read more exactly $+c_y e_Y^i$ in the notation of \cite{Bini:2026dvn}.]
  The same contribution is contained both in the MPM waveform 
and in the 
EFT one before adding in the latter the effect of the Veneziano-Vilkovisky supertranslation (which, in fact, doubles a term
already present in the EFT waveform). To complete these  remarks, let us also note 
that the 1PM-level definition of $J_{\mu \nu}^{\rm sys}$
(containing explicit $O(G)$ field contribution) in \cite{Bini:2022wrq} does not contain any time-odd contribution.

\section{Comparison with soft theorems}

Let us finally briefly discuss the comparison with soft theorems \cite{Saha:2019tub,Sahoo:2021ctw,Sen:2024qzb}.
This comparison is important because it yields checks of our results up to  the 2-loop level included (corresponding to the
terms of order $G^3$ in $U_3$).

In the soft limit ($\omega \to 0$, i.e., $u\to 0$) we find for our 2-loop, MPM-derived expression of $U_3$ an expansion whose first two terms read
\bea
\label{U3soft}
\hat U^{\rm soft}_{3,\,\rm MPM}&=& i\, \frac{{\mathcal A}_3}{u}+{\mathcal B}_3 \ln u+O(u\ln^2 u)\,,
\eea
where  
\bea
{\mathcal A}_3&=& {\mathcal A}^{G^1}+{\mathcal A}^{G^2}+{\mathcal A}^{G^3}\,,\nonumber\\
{\mathcal B}_3&=& {\mathcal B}^{G^1}+{\mathcal B}^{G^2}+{\mathcal B}^{G^3}\,.
\eea
Explicitly, we find
\bea
\label{our_res}
{\mathcal A}^{G^1}&=& \nu X_{12} GM^2 \bar m_2 (\bar m_2n_1+2\bar m_1 n_2)  {\sf a}^{G^1}(p_\infty,\nu)\nonumber\\
{\mathcal A}^{G^2}&=& \nu X_{12} \frac{G^2M^3}{b} \bar m_2 (\bar m_2n_1+2\bar m_1 n_2) \pi  {\sf a}_\pi^{G^2}(p_\infty,\nu)\nonumber\\
{\mathcal A}^{G^3}&=& \nu X_{12} \frac{G^3M^4}{b^2} \left[\frac{{\sf a}^{G^3}_{-4}}{p_\infty^4}+\frac{{\sf a}^{G^3}_{-2}}{p_\infty^2}+{\sf a}^{G^3}_{0}\right.\nonumber\\
&+&\left. (\pi {\sf a}^{G^3}_{1,\pi}+{\sf a}^{G^3}_{1,\slashed{\pi}})p_\infty+{\sf a}^{G^3}_{2}p_\infty^2\right] \,,
\eea
with
\bea
{\sf a}^{G^1}(p_\infty,\nu)&=& 2+\frac23 (\nu+4)p_\infty^2+\left(-\frac{56}{33}+\frac{41}{66}\nu-\frac{14}{33}\nu^2 \right)p_\infty^4\nonumber\\
&+& \left(\frac{512}{429}-\frac{443}{572}\nu-\frac{145}{429}\nu^2 +\frac{10}{39}\nu^3 \right)p_\infty^6\,,\nonumber\\
{\sf a}_\pi^{G^2}(p_\infty,\nu)&=&3+\left(\frac74+\nu \right)p_\infty^2
+\frac{1}{22}(-23+4\nu-14\nu^2)p_\infty^4\,,\nonumber\\
\eea
and
\bea
\label{contr_G3}
{\sf a}^{G^3}_{-4}&=& 2\bar m_1^2 n_1 -3\bar m_2 (\bar m_2n_1+2\bar m_1 n_2) \,,\nonumber\\
{\sf a}^{G^3}_{-2}&=& \frac83 (4+\nu)\bar m_1^2 n_1-4(1+\nu)\bar m_2(\bar m_2n_1+2\bar m_1 n_2)\,,  \nonumber\\
{\sf a}^{G^3}_{0}&=& \frac{4}{33}n_1 \bar m_1^2 (140+111\nu +2\nu^2)-\frac{2}{11}\bar m_2 (\bar m_2n_1+2\bar m_1 n_2)\,,\nonumber\\
{\sf a}^{G^3}_{1,\pi}&=&\nu\left[ \frac{25309}{2880}\bar m_2^2 n_2+\frac{409}{960}\bar m_1 (\bar m_1 n_2+2\bar m_2 n_1)\right]\,,\nonumber\\
{\sf a}^{G^3}_{1,\slashed{\pi}}&=&\frac{16}{5}\nu \bar m_2 (\bar m_2n_1+2\bar m_1 n_2)\,, \nonumber\\
{\sf a}^{G^3}_{2}&=& \frac{1}{429} \bar m_1^2 n_1 (2176+8121\nu+466\nu^2-72\nu^3)\nonumber\\
&+& \frac{\bar m_2 }{858}(\bar m_2n_1+2\bar m_1 n_2)(7616-26027 \nu-4726\nu^2+216 \nu^3)\,.\nonumber\\
\eea
Defining $S_{3m}$ and $D_{3m}$ (for $m=3,1,\bar 1,\bar 3$) as,
\bea
S_{3m}&\equiv &{}_{-2}Y_{3m}+{}_{-2}Y_{3\bar m}\,,\nonumber\\
D_{3m}&\equiv &{}_{-2}Y_{3m}-{}_{-2}Y_{3\bar m}\,,
\eea
the results above involve the following polarization structures
\bea
X&\equiv & \bar m_2 (\bar m_2n_1+2\bar m_1 n_2)\nonumber\\
&=&  \frac{1}{5}\sqrt{\frac{\pi}{14}}  (\sqrt{5} D_{31} + 5 \sqrt{3} D_{33}),\nonumber\\ 
Y&\equiv & \bar m_1^2 n_1\nonumber\\
&=& \frac{1}{15} \sqrt{\frac{\pi}{14}} (3\sqrt{5} D_{31} - 5 \sqrt{3}D_{33}), 
\eea
except for ${\sf a}^{G^3}_{1,\pi}$ which involves  the  $1\leftrightarrow 2$ transforms of $X$ and $Y$, namely  
\bea
{\widetilde X}&\equiv & \bar m_1 (\bar m_1n_2+2\bar m_2 n_1)\nonumber\\
&=& -\frac{i}{5} \sqrt{\frac{\pi}{14}}(\sqrt{5} S_{31} - 5 \sqrt{3} S_{33}), \nonumber\\
{\widetilde Y}&\equiv & \bar m_2^2 n_2\nonumber\\
&=& -\frac{i}{15} \sqrt{\frac{\pi}{14}}(3 \sqrt{5} S_{31} + 5 \sqrt{3} S_{33})\,.
\eea

All  the contributions \eqref{our_res} have been checked against the general soft limit expressions obtained in Refs. \cite{Saha:2019tub,Sahoo:2021ctw,Sen:2024qzb} which yield 
\bea
\label{ABsen}
{\mathcal A}&=& \frac{b}{p_\infty}\check A_{\mu\nu}\bar m^\mu \bar m^\nu\,,\nonumber\\
{\mathcal B}&=& (\check C_{\mu\nu}-\check B_{\mu\nu})\bar m^\mu \bar m^\nu\,,
\eea
where $\check A_{\mu\nu}$, etc. differs from $A_{\mu\nu}$, etc. in Ref. \cite{Sen:2024qzb} by discarding a prefactor $2G/(c^4 R)$.
Let us recall the subtleties about the contributions of massless particles (i.e. emitted gravitons) in the outgoing state.

In the ${\mathcal A}$ coefficient they must be explicitly added and they correspond to the nonlinear memory. It happens that
we have evaluated the octupolar part of the nonlinear memory in our recent work \cite{Bini:2026vaq}, namely (when truncating
it at the leading PM order, i.e.,  $O(G^3)$ (2-loop), and at the 2.5PN order): 
\bea
\label{nonlinmem}
{\mathcal A}^{G^3{\rm nl\, mem}}_{l=3}&=& -\frac{G^3 M^4 \nu^2 \pi   X_{12}}{b^2}\left( \frac{409}{960}  {\widetilde X}  +
 \frac{3997}{2880}   {\widetilde Y}  \right)p_\infty\,.\nonumber\\
\eea
Here we took into account that in Ref. \cite{Bini:2026vaq} $\sqrt{1-4\nu}$ denotes $X_{21}=-X_{12}$.

In the ${\mathcal B}$ logarithmic coefficient the contribution of massless particles to $W= h/(4 G)$ is equal to $-2G(P_F^\mu P_F^\nu - P_I ^\mu P_I^\nu)\bar m_\mu \bar m_\nu$ (see Eq. (2.13) in Ref. \cite{Sen:2024qzb}).  However, the difference between the
total outgoing momentum carried by massive particles $P_F^\mu$ and the total incoming momentum carried by massive particles $P_I ^\mu$ is of order $G^3$. After multiplication by $G$, the contribution of this term to the logarithmic ${\mathcal B}$ 
contribution to $W^{\rm soft}$ is of order $G^4$, i.e. at the 3-loop level. We can therefore neglect it here.

In addition to the direct occurrence of massless graviton contributions in the soft expansion, it is crucial
to take into account the indirect effect of radiation graviton exchange between the massive particles, i.e. of
radiation-reaction effects. In the framework of Ref. \cite{Bini:2021gat}, radiative contributions to the outgoing momenta
start at order $\Delta^{\rm rr} p_a^\mu =O(G^3)$ and are given by the sum of three different effects:
(i) the inclusion of a radiative contribution $\chi^{\rm rr}$ to the {\it relative} scattering angle; (ii) the effect
(considered in the cm system) of the total radiated energy $E_{\rm rad}=P^0_{\rm rad}$; and (iii) the effect of the
cm-recoil, linked to the total radiated momentum, which has components both in the $x$ and $y$ directions:
$P^x_{\rm rad}$ and $P^y_{\rm rad}$.  While $P^x_{\rm rad}=O(G^4)$ can be neglected here, we a priori
need to take into account the effect of $P^y_{\rm rad}=O(G^3)$ on the outgoing momenta of the binary system.
We found that the 2-loop contribution linked to $P^y_{\rm rad}$ does not appear at 2.5PN.

The conservative contribution to the scattering momenta reproduce all the above displayed MPM values of ${\mathcal A}$, apart from the $O(G^3\, p_\infty)$ terms   in  ${\sf a}^{G^3}_{1,\pi}$ and ${\sf a}^{G^3}_{1,\slashed{\pi}}$, Eq. \eqref{contr_G3}.
We then found that the non-$\pi$ contribution ${\sf a}^{G^3}_{1,\slashed{\pi}}$ is fully explained by the effect of 
\beq
\chi_{\rm rr}^{\rm LO}=\frac{16}{5}\frac{\nu}{p_\infty} \frac{G^3M^3}{b^3}\,,
\eeq
while the $\pi$ contribution ${\sf a}^{G^3}_{1,\pi}$ is equal to the sum of the nonlinear memory term, Eq. \eqref{nonlinmem} and of the term 
$\delta {\mathcal A}=\frac{37}{5} \bar m_2^2  n_2 \pi p_\infty$ 
sourced by the LO radiated energy
\beq
E_{\rm rad}^{\rm LO}=\frac{37}{15}\pi \nu^2 \frac{G^3 M^4}{b^3} p_\infty \,.
\eeq

Let us consider now the next term in the soft limit expansion of $U_3$: ${\mathcal B}_3$, Eq. \eqref{U3soft}.
We find
\begin{widetext}
\bea
{\mathcal B}^{G^1}&=& GM^2 X_{12}\nu \bar m_2^2n_2 \left[2+\left(-\frac{13}{3}+\frac{2\nu}{3}  \right)p_\infty^2
+\left(-\frac{14 \nu ^2}{33}-\frac{113 \nu }{66}+\frac{95}{132}\right)p_\infty^4
+\left( \frac{10 \nu ^3}{39}+\frac{164 \nu ^2}{143}+\frac{151 \nu }{156}-\frac{259}{3432}\right)p_\infty^6   \right]\,,\nonumber\\
{\mathcal B}^{G^2}&=& G^2M^3 \frac{X_{12}\nu}{b} \bar m_2 (\bar m_2n_1+2\bar m_1 n_2)\left[-4p_\infty-\frac23 (8+5\nu)p_\infty^3 \right]\,,  \nonumber\\
{\mathcal B}^{G^3}&=& G^3M^4 \frac{X_{12}\nu}{b^2}[2 \bar m_1 (2  \bar m_2 n_1+\bar m_1 n_2)-3\bar m_2^2 n_2]\times\nonumber\\
&&\left[\frac{1}{p_\infty^4}+\frac{11+8\nu}{6p_\infty^2}
+\left(\frac{4 \nu ^2}{33}+\frac{68 \nu }{33}-\frac{379}{88}\right)
+\left(-\frac{12 \nu ^3}{143}+\frac{17 \nu ^2}{143}-\frac{2680 \nu }{429}-\frac{49867}{6864}\right)p_\infty^2\right]\nonumber\\
&-& G^3 M^4 \frac{X_{12}\nu}{b^2} 6\pi \bar m_2 (\bar m_2 n_1 +2\bar m_1 n_2)p_\infty\,.
\eea
\end{widetext}
We have checked that all the above contributions agree  with the results of inserting the exact soft expressions  of Ref. \cite{Sen:2024qzb}  in Eq. \eqref{ABsen}. We notice that, contrary to the case of ${\mathcal A}$,  the radiative contributions to the impulses start  contributing to ${\mathcal B}$ at order $G^4$ and   therefore do not contribute to the present 2-loop ${\mathcal B}$-comparison.

\section{Concluding remarks}

We have extended the accuracy (both in the time-domain, and in the frequency-domain)
of the computation of  the gravitational waveform emitted during the scattering of two masses by evaluating the even-parity octupolar waveform
to the  2-loop level  ($O(G^4)$ in ${\bar m}^\mu {\bar m}^\mu h_{\mu \nu}$) and to the
absolute 3.5PN accuracy (i.e. the fractional 3PN accuracy in the radiative octupole moment $U_{ijk}$). 
At this accuracy,  the waveform includes contributions linked to  the 2.5PN ($G^2+G^3$) radiation-reacted hyperbolic motion.

We presented several  checks of our results: logarithmic structure, soft limit, first order self-force limit and  comparisons with the EFT 1-loop waveform. All those tests were successful. 
In particular, the comparison with the EFT waveform showed coincidence only  
when  taking into account exactly the {\it same} (2.5PN-level) difference in the definitions of the 
 position of the center-of-mass  within the MPM and EFT formalisms that was deduced from our previous comparison
 of the radiative quadrupole at the (absolute and fractional) 3.5PN-level. The latter difference 
 is equal to 
\beq
X_{\rm EFT \, cm}^i - X_{\rm MPM \, cm}^i = - c_y e_y^i\,,
\eeq
where 
\beq
\label{cynew}
c_y=+\frac45 \frac{G }{c^5}\frac{m_1 m_2 (m_1-m_2)}{(m_1+m_2)^2}p_\infty^3\,.
\eeq

 The full effect on the total waveform $W$
 of the difference $X_{\rm EFT \, cm}^i - X_{\rm MPM \, cm}^i$
 in the center-of-mass origins is 
\beq
 \delta W \equiv W^{\rm EFT} - W^{\rm MPM}=+ i \eta \omega (c_y n^y )W\,,
\eeq
where $c_y =O(G \eta^5)$ is given by Eq. \eqref{cynew}, and where $W$ has both a PM expansion, $W\sim G^1+G^2+G^3+\cdots$ and a PN one (which includes the $\eta^n$ prefactors indicated in Eq. \eqref{W_deco}).
When evaluating $W$ at the absolute $\eta^7$ (3.5PN) accuracy and comparing $W^{\rm MPM}$ to the one-loop
waveform the only terms we need to worry about come from
\beq
\delta  W=i \eta \omega (c_y n^y ) ( U_2^{\rm tree}+ \eta (V_2^{\rm tree} +U_3^{\rm tree}) )\,,
\eeq
We leave to future studies  an extension of the present work to other multipolar components of the bremsstrahlung waveform,
and notably to the odd-parity sector. The latter sector should include the term $i \eta \omega (c_y n^y )\eta V_2^{\rm tree}$, which will contribute to $V_3$ at the fractional 2.5PN level.\\

\section*{Acknowledgments}

D.B. acknowledges membership to the Italian Gruppo Nazionale per
la Fisica Matematica (GNFM) of the Istituto Nazionale
di Alta Matematica (INDAM), as well as the hospitality
and the highly stimulating environment of the Institut
des Hautes Etudes Scient\'if\'iques.
A.G.  is grateful to the Istituto
per le Applicazioni del Calcolo \lq\lq M. Picone," CNR, Rome (IT) for
past support and hospitality during the development of
the present project.
The present research was partially supported by the
2021 Balzan Prize for Gravitation: Physical and Astrophysical Aspects, awarded to T. Damour.

\end{document}